\documentclass[
 reprint,
 superscriptaddress,
 amsmath,amssymb,
 aps,
 prl
]{revtex4-2}

\usepackage{graphicx}
\usepackage{dcolumn}
\usepackage{bm}
\usepackage[hidelinks,colorlinks=true,linkcolor=blue,citecolor=blue, anchorcolor = blue, urlcolor = blue]{hyperref}
\usepackage[mathlines]{lineno}

\usepackage{braket}
\usepackage{color}
\usepackage{makecell}

\newcommand{\affsiqse}{\affiliation{Shenzhen Institute for Quantum Science and Engineering (SIQSE), Southern University of Science and Technology, Shenzhen, P. R. China.}}
\newcommand{\affiqa}{\affiliation{International Quantum Academy, Shenzhen 518048, China.}}
\newcommand{\affgdlab}{\affiliation{Guangdong Provincial Key Laboratory of Quantum Science and Engineering, Southern University of Science and Technology Shenzhen, 518055, China.}}

\newcommand{\yb}{^{171} \mathrm{Yb}^+}
\newcommand{\figref}[2]{Fig.~\ref{#1}\textcolor{blue}{#2}}

\newcommand{\tabref}[1]{Tab.~\ref{#1}}

\begin{document}

\title{
  High-Fidelity Detection on $\yb$ Qubit via $^2D_{3/2}$ Shelving
}

\affsiqse
\affiqa
\affgdlab

\author{Xueying Mai}
  \thanks{These two authors contributed equally to this work.}

\author{Liyun Zhang}
  \thanks{These two authors contributed equally to this work.}
  
\author{Yao Lu}
  \email{luy7@sustech.edu.cn}

\date{\today}

\begin{abstract}
High-fidelity detection of quantum states is indispensable for implementing quantum error correction, a prerequisite for fault-tolerant quantum computation. For promising trapped ion qubits, however, the detection fidelity is inherently limited by state leakage. Here, we propose an efficient approach to enhance the fidelity of detecting $\yb$ qubits through $^2D_{3/2}$ state shelving techniques. Leveraging selective shelving and state-dependent fluorescence, we mitigate the impact of state leakage and experimentally realize a fidelity of 99.88(2)\%, while over 99.99\% fidelity is predicted by utilizing state-of-the-art hardwares. Meanwhile, we demonstrate the feasibility of mid-circuit measurements, a crucial step for recent implementations of quantum error correction, by mapping the hyperfine qubit to metastable levels. Our research provides an essential component for realizing fault-tolerant quantum information processing with trapped-ion systems in the near future.

\end{abstract}

\maketitle

\section{
  \label{sec:intro}
  Introduction
}

Among the diverse physical platforms available for quantum information processing, trapped ions have emerged as a leading candidate for realizing large-scale quatnum computation \cite{kielpinski2002architecture, monroe2013scaling, brown2016co}. Qubits encoded in individual ions exhibit exceptionally long coherence time \cite{wang2017single,wang2021single}. High-fidelity quantum logic gates have been well demonstrated on single, two or even multiple ion qubits \cite{PhysRevLett.131.120601, PhysRevLett.117.060504, PhysRevLett.117.060505, PhysRevLett.127.130505, srinivas2021high, lu2019global, figgatt2019parallel}. Beyond the realm of coherent operations, the accuracy of qubit state detection also stands as a cruicial metric for evaluating quantum processors \cite{divincenzo2000physical}. High-fidelity detection plays a pivotal role in enabling quantum error correction and reducing the required resources \cite{egan2021fault, postler2022demonstration, PhysRevX.13.041057}. Therefore, in the quest to unlock the immense computational power offered by quatnum processors, pursing high-fidelity detection on qubits represents one of the critical research frontiers.

In trapped-ion systems, a single ion qubit can be measured by cyclically exciting only one of the qubit state (refer to "bright" state) to short-live auxiliary levels. The spontaneously emitted photons during the cycling transition are collected to determine whether the qubit is projected to the bright or, conversely, the "dark" state \cite{wineland1995quantum}. However, off-resonant couplings between undesired levels can cause qubit state leakage during the detection process, representing one of the main errors in measuring ion qubits. For favored $\yb$ ions \cite{PhysRevA.76.052314}, the leakage error can be substantial, given that the energy gap of the qubit encoded in the ground-state hyperfine structure is only several GHz. Consequently, the detection fidelity for single $\yb$ qubit was once limited to around 99\% \cite{PhysRevA.76.052314, PhysRevA.82.063419}.

Various efforts have been made to overcome such limitations, with one representive approach focusing on improving the efficiency of photon collection and subsequently reducing the detection duration. This direction has been greatyly pursed by employing objective lenses with high numerical aperature (N.A.) and high quantum effeciency (QE) single photon detectors \cite{noek2013high, crain2019high}. Additionally, inspired by high-fidelity detection for optical ion qubits \cite{PhysRevLett.113.220501, keselman2011high}, the shelving method has been introduced for $\yb$ qubits, utilizing either $^2D_{5/2}$ or even $^2F_{7/2}$ manifolds \cite{PhysRevA.104.012606}. However, the lifetime of $^2D_{5/2}$ is comparatively short, and the state transfer between $^2S_{1/2}$ and $^2F_{7/2}$ is not that efficient.

In this work, we propose and experimentally demonstrate a novel approach for improving the detection fidelity on single $\yb$ ion qubit, by selectively shelving a portion of the qubit state to the $^2D_{3/2}$ manifold. Our results successfully show a significant reduction in errors attributed to qubit-state leakage. In our current experimental configuration, we achieve a final detection fidelity of 99.88(2)\%, primarily limited by the photon collection efficiency which is approximately 0.763(4)\%. Leveraging state-of-the-art hardwares and compatible tools \cite{crain2019high, PhysRevApplied.12.014038, seif2018machine, jeong2023using}, we estimate that the detection fidelity can surpass 99.99\%. Furthermore, by shelving the entire qubit onto a pair of Zeeman levels within $^2D_{3/2}$, we also demonstrate the feasibility of applying mid-circuit measurements \cite{allcock2021omg}. Our research provide a novel pathway for achieving highly accurate detection of $\yb$ ion qubits and contributes to the realization of quantum error correction in the near future.

\section{
  \label{sec:scheme}
  Scheme and setup
}

\begin{figure*}[htbp]
  \includegraphics[scale = 1]{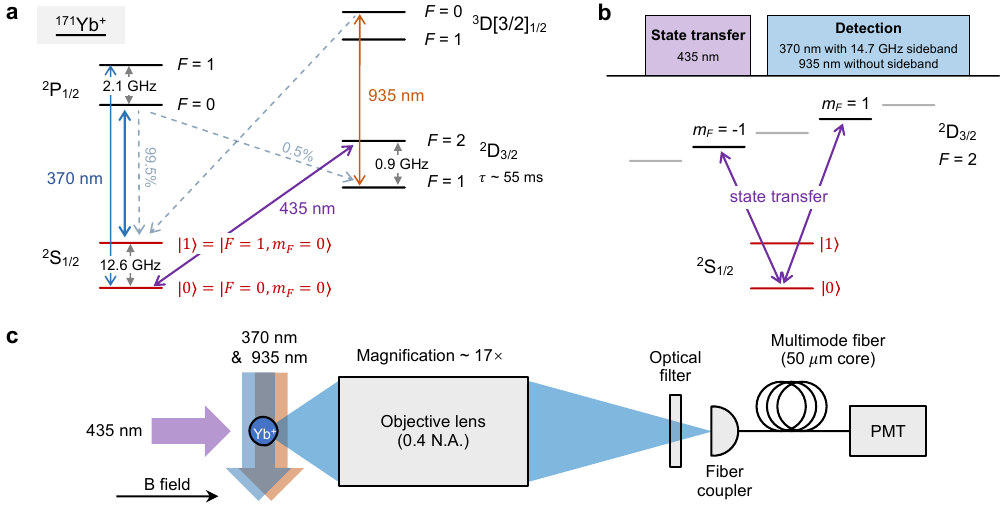}
  \caption{
  \label{fig:setup}
  Schemes for $\yb$ qubit detection and experimental setup.
  \textbf{a}  Relative energy levels of a single $\yb$ ion. Relevant transition wavelengths used in the experiment are labeled with colored double-arrow lines. The decay channels of the $^2P_{1/2}$ manifold are marked as well (dashed line).
  \textbf{b} Scheme and sequence for $^2D_{3/2}$ shelving detection. The state in the qubit $\ket{0}$ level is coherently transferred to the $^2D_{3/2}\ket{F=2}$ levels. In the experiment, both $^2D_{3/2}\ket{F=2, m_F = \pm1}$ are utilized for a two-step transfer to mitigate the shelving error.
  \textbf{c} Experimental setup.
  The laser beams of 370~nm and 935~nm are co-aligned and directed onto the ion, while the 435~nm laser is aligned along the quantization axis. A magnetic field of approximately 10~Gauss is applied in our system to degenerate all Zeeman levels. The fluorescence emitted by the ion is imaged by an objective lens (0.4 N.A.) with a magnification round 17, and then delivered to a single photon detector through a multimode fiber (core diamater 50~$\mu$m). 
  }
\end{figure*}

A individual qubit for $\yb$ is typically encoded in the hyperfine levels belonging to the ground manifold, denoted as $\ket{0} \equiv {}^2S_{1/2}\ket{F = 0, m_F = 0}$ and $\ket{1} \equiv {}^2S_{1/2}\ket{F = 1, m_F = 0}$, as shown in \figref{fig:setup}{a}, and its energy gap is around 12.6~GHz \cite{PhysRevA.76.052314}. For the standard approach to discriminate the qubit state in the hyperfine levels (marked as the hyperfine detection here and below), we directly apply a laser field of 370~nm which is resonant to the cycling transition $^2S_{1/2}\ket{F = 1} \leftrightarrow  {^2P_{1/2}\ket{F = 0}}$, and consequently a lot of photons would be scattered is the $\ket{1}$ state is projected. Moreover, an additional laser of 935~nm is simultaneously applied to repump the ion state from $^2D_{3/2}\ket{F = 1}$ back to the cycle. 

During the hyperfine detection process, there is a chance for the bright $\ket{1}$ (dark $\ket{0}$) state to be off-resonantly excited to the $^2P_{1/2}\ket{F = 1}$, leading to a potential decay to the dark $\ket{0}$ (bright $\ket{1}$) state. Such state leakage can significantly impact detection fidelity \cite{noek2013high, crain2019high}. To mitigate this issue, one strategy is to transfer a part of the qubit state to another atomic level which is far off-resonant from the cycling transition, known as the shelving techniques. Both the $^2D_{5/2}$ and $^2F_{7/2}$ manifolds have been employed in shelving detection \cite{PhysRevA.104.012606}. However, the $^2D_{5/2}$ manifold has a relatively short lifetime. In the case of the $^2F_{7/2}$ state, the octupole transition directly from $^2S_{1/2}$ has a limited speed, or alternatively, a two-step state transfer must be employed to address this limitation \cite{yang2022realizing}.

In addition to the aforementioned options, it's worth noting that the hyperfine structure $^2D_{3/2}\ket{F = 2}$ can also serve as a viable choice for state shelving. These levels do not participate in the detection process, and their lifetime is comparatively long (around 55~ms \cite{PhysRevResearch.5.023193}). \figref{fig:setup}{b} provide a brief illustration of our shelving detection scheme (marked as shelving detection here and below). In detail, during the shelving detection stage, the state occupation in the $\ket{0}$ state is first transferred to the $^2D_{3/2}\ket{F = 2}$ levels via an electric quadrupole transition, then resonant cycling and repump transitions are driven as the hyperfine detection. Note that other than 935~nm laser, the transition from $^2S_{1/2}\ket{F = 0}$ to ${^2P_{1/2}\ket{F = 1}}$ is driven simultaneously in our shelving scheme, to pump out the state that leaks to the $^2S_{1/2}\ket{F = 0}$ level.

In \figref{fig:setup}{c}, we briefly layout our experimental setup that utilized to examine our detection scheme. A single $\yb$ ion is trapped in a segmented blade trap (not shown in the figure), and the ion's fluorescence is collected by an objective lens with a numerical aperture (N. A.) of 0.4. The collected photons pass through a narrow-band optical filter and are subsequently coupled to a multimode fiber, effectively suppressing noise from background scattering. The collected photons are then delivered to a commercial photomultiplier tube (PMT) for photon counting. The total collection efficiency of a single photon is approximately 0.763(4)\%, with details on the contributions from each component provided in the Methods section. The 435~nm laser required for state shelving is stabilized to a high-finesse cavity to reduce its linewidth to below 15~Hz, while the 370~nm laser for the cycling transition and the 935~nm laser for state repump are locked to a wavelength meter.

\section{
  \label{sec:results}
  Results
}

\subsection{Characterization of state leakage rates}

Since the primary source of infidelity in ion qubit measurement arises from state leakage induced by detection-related laser fields, we first analyze the efficacy of our shelving detection approach in suppressing such rates. Specifically, for the qubit state $\ket{1}$, the dark states during shelving detection manifest as levels belong to $^2D_{3/2}\ket{F = 2}$. Leakages to these states can be traced to the route $^2S_{1/2}\ket{F = 1} \rightarrow {^2P_{1/2}\ket{F = 1}} \rightarrow {^2D_{3/ 2}\ket{F = 2}}$. The first transition, $^2S_{1/2}\ket{F = 1} \rightarrow {^2P_{1/2}\ket{F = 1}}$, is induced by off-resonant coupling via the 370~nm laser, while the subsequent step results from spontaneous emission. Despite the off-resonant coupling rate aligning with the hyperfine detection method, the overall leakage rate is mitigated due to the relatively small branching ratio of spontaneous emission from ${^2P_{1/2}\ket{F = 1}}$ to ${^2D_{3/ 2}\ket{F = 2}}$, estimated to be approximately 0.5\%. Consequently, we estimate that the leakage rate for the $\ket{1}$ state can be suppressed by approximately two orders of magnitude compared to the direct hyperfine detection. For the qubit state $\ket{0}$, one leakage channel is $^2D_{3/2}\ket{F = 2} \rightarrow {^3D[3/2]_{1/2}\ket{F = 1}} \rightarrow {^2S_{1/ 2}\ket{F = 1~\mathrm{or}~0}}$, wherein the first transition is induced by the off-resonant coupling via the 935~nm laser.

\begin{figure}[htbp]
  \includegraphics[scale = 1.0]{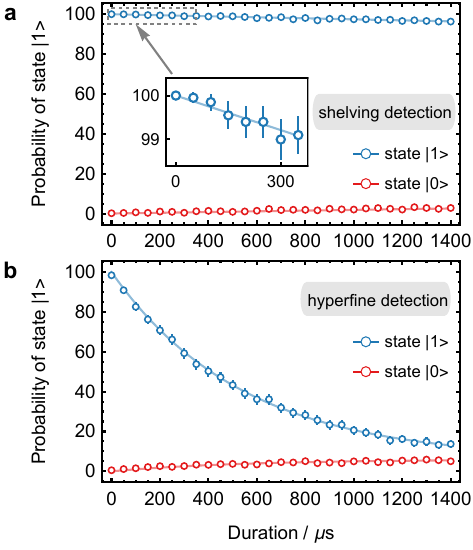}
  \caption{
    \label{fig:leakage_rates}
    Experimental results for measuring leakage rates.
    After preparing the qubit to either $\ket{1}$ (blue) or $\ket{0}$ (red) state, detection-related lasers are intentionally applied to induce state leakage, and the final state is measured versus the duration of applied lasers.
    Subfigures \textbf{a} and \textbf{b} respectively depict the results for shelving detection and hyperfine detection. The circular markers represent the experimental data, while the solid lines denote the fitting results used to extract the flip rates. All error bars presented here and below represent 95\% confidence intervals.
  }
\end{figure}

We experimentally characterize the leakage rates of our $^2D_{3/2}$ shelving detection and compare them with the hyperfine detection approach. The results are illustrated in \figref{fig:leakage_rates}. Here, we initialize the qubit to either the $\ket{1}$ or $\ket{0}$ state, followed by the utilization of detection-related lasers to induce leakage with varied durations. The measurement is applied at the end to record the final qubit state. As expected, we can clearly observe a reduction in the leakage rate for the case of the initial $\ket{1}$ state. The fitted results extracted from summarized in the first row of \tabref{tab:flipping-rates} reveal a $67^{+9}_{-8}$ times suppression for the rate $R_{\ket{1}\rightarrow\ket{0}}$, while the theoretical prediction is around 53 times (see Methods section).

\begin{table}[htbp]
  \centering
  \tabcolsep = 0.2cm
  \renewcommand\arraystretch{1.5}
  \caption{\label{tab:flipping-rates}%
  Leakage rates comparison
  }
    \begin{tabular}{c|c|c}
      \hline\hline
      Detection scheme& \thead{Hyperfine} & \thead{$^2D_{3/2}$ shelving} \\
      \hline
      $R_{\ket{1} \rightarrow \ket{0}}$ ($\mu \mathrm{s}^{-1}$) & $17.4(6)\times10^{-4}$ & $2.6(2)\times10^{-5}$ \\
      $R_{\ket{0} \rightarrow \ket{1}}$ ($\mu \mathrm{s}^{-1}$) & $1.1(3)\times10^{-4}$ & $1.9(3)\times10^{-5}$\\
      \hline
      \hline
    \end{tabular}
\end{table}

\begin{figure}[htbp]
  \includegraphics[scale = 1.0]{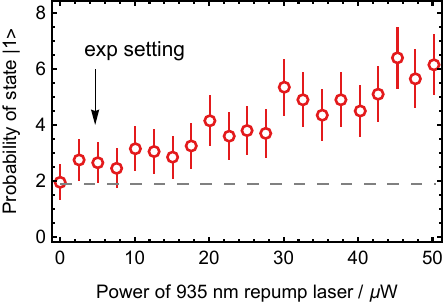}
  \caption{
    \label{fig:935power_dependence}
    State leakage for the $\ket{0}$ state induced by the off-resonant coupling from 935~nm laser.
    After initializing the qubit to the $\ket{0}$ state, we apply the detection-related lasers at a fixed time of 1~ms to induce the state leakage. The power of the 935~nm laser is varied to characterize the flip induced by the off-resonant coupling. As expected, we observe an increase in the probability of detecting state $\ket{1}$ with increasing power of the 935~nm laser. The gray dashed line indicates the lower bound limited by the lifetime of the $^2D_{3/2}$ levels. Experimentally, we set the 935~nm laser power to be around 4.7~$\mu$W (beam diamater around 87 $\mu$m), which is high enough to pump out the state leaking to $^2D_{3/2}\ket{F = 1}$.
  }
\end{figure}

However, the suppression of the leakage rate $R_{\ket{0}\rightarrow\ket{1}}$ via $^2D_{3/2}$ shelving is limited, as displayed in the second row of \tabref{tab:flipping-rates}. The main reason for this limitation is the lifetime of the $^2D_{3/2}$ levels ($\tau_{D_{3/2}}\sim$55 ms \cite{PhysRevResearch.5.023193}), which bounds the $R_{\ket{0}\rightarrow\ket{1}}$ to around $1/\tau_{D_{3/2}}=1.8\times10^{-5}~\mu\mathrm{s}^{-1}$. This value is consistent with our experimentally achieved minimal $R_{\ket{0}\rightarrow\ket{1}}$ of $1.9(3)\times10^{-5}$ $\mu\mathrm{s}^{-1}$. Above results also indicate that the error caused by the off-resonant coupling from the 935~nm laser is negligible. We also investigate the increase in $R_{\ket{0}\rightarrow\ket{1}}$ by varying the power of the 935~nm laser, as depicted in \figref{fig:935power_dependence}.

\subsection{High-fidelity detection via $^2D_{3/2}$ shelving}

\begin{figure}[htbp]
  \includegraphics[scale = 1.0]{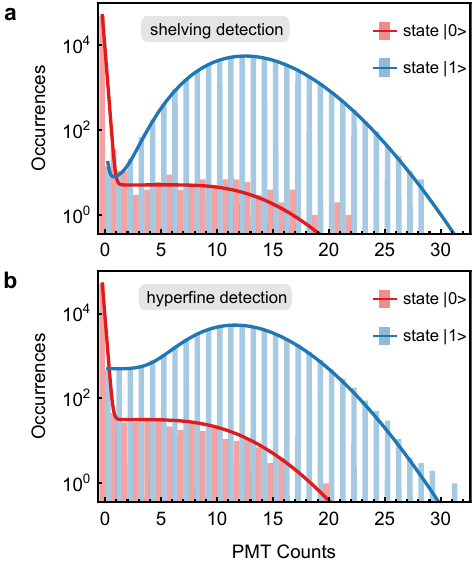}
  \caption{\label{fig:photon_histogram}
  Histogram of the number of photons collected by utilizing \textbf{a} $^2D_{3/2}$ shelving detection and \textbf{b} hyperfine detection. For both $\ket{0}$ and $\ket{1}$ states we repeat experiment for $5\times10^{4}$ times. The solid lines in each subfigure represent the theoretical distribution by utilizing leakage rates in \tabref{tab:flipping-rates} \cite{acton2005near}. For shelving detection, the errors for detecting $\ket{0}$ state ($\epsilon_0$) and $\ket{1}$ state ($\epsilon_1$) are $1.8(4)\times10^{-3}$ and $0.5(2)\times10^{-3}$, respectively, by setting threshold to 1. These errors increase to $\epsilon_0=6.4(7)\times10^{-3}$ and $\epsilon_1=9.3(8)\times10^{-3}$ for hyperfine detection, by setting threshold to 0. The values of the average photon number for both detection schemes are around 12.}
\end{figure}

After evidently demonstraing the suppressing of leakage rates by utilizing $^2D_{3/2}$ shelving approach, we further verify its capability in improving the detection fidelity. The results with optimized experimental parameters are summarized in \figref{fig:photon_histogram}. Here, we fix the detection duration to 70~$\mu$s, and for each prepared qubit state, we repeat the experiments $5\times10^4$ times.

Ideally, distributions of the collected photons follow the Poisson distribution. However, leakage events that occur during the detection process can cause the bright (dark) state to have undesired tail distributions at low (high) photon counts. Comparing the histograms shown in \figref{fig:photon_histogram}{a} and \textcolor{blue}{b}, we can clearly observe that these undesired tail are suppressed by utilizing $^2D_{3/2}$ shelving. By simply employing the threshold discrimination, we estimate the average detection error for shelving detection to be $\epsilon = 1.2(2)\times10^{-3}$ (99.88(2)\% fidelity), 
and the theoretical prediction is $0.7\times10^{-3}$.
While for the hyperfine detection, the average error increase to $\epsilon = 7.8(6)\times10^{-3}$ (99.22(6)\% fidelity).

Except for the intrinct error due to the state leakage, addtional errors in our experimental demonstration arise from the state preparation and the shelving stage. With independent benchmarking, the error of the microwave $\pi$-rotation to prepare the $\ket{1}$ state is around $4\times10^{-4}$, while that of the 435~nm $\pi$-rotation to state transfer is around $10^{-2}$. For the latter error, we mitigate it by applying two-step state transfer (as shown in \figref{fig:setup}{b}), suppressing the overall shelving error to $10^{-4}$ level.

\begin{figure}[htbp]
  \includegraphics[scale = 1.0]{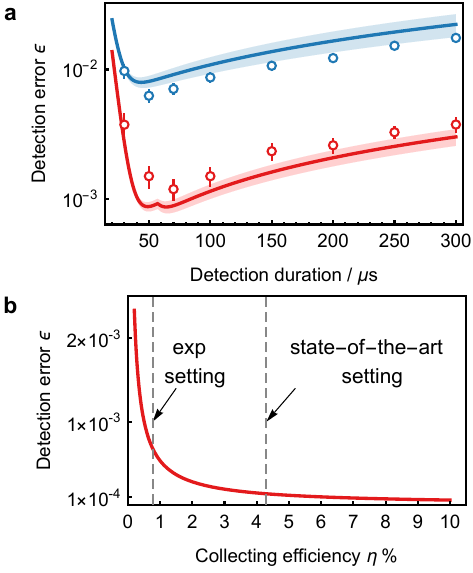}
  \caption{
    \label{fig:error_scaling}
    Detection error versus detection duration and photon collection efficiency. In \textbf{a}, experimental data for shelving detection (red) and hyperfine detection (blue) are represented by circle markers. The solid lines in both \textbf{a} and \textbf{b} correspond to theoretical predictions derived from the leakage rates presented in \tabref{tab:flipping-rates}.
    The dashed lines displayed in \textbf{b} indicate our current collection efficiency and the state-of-the-art record among previous works.
  }
\end{figure}

In \figref{fig:error_scaling}{a}, we illustrate the scaling of the detection error versus detection duration. The detection error reaches its minimum under similar detection durations for both shelving and hyperfine detection schemes. Unlike the $^2D_{5/2}$ shelving scheme utilized in calcium ion qubits \cite{PhysRevLett.113.220501}, the $^2D_{3/2}$ shelving method employed here cannot entirely eliminate state leakage. Therefore, we cannot reduce the detection error by simply extending the detection time.

To further suppress the detection error, an increase in photon collection efficiency is still required. Utilizing the state leakage rates obtained in our experiments, we ideally estimate the scaling of the detection error versus the collection efficiency, as shown in \figref{fig:error_scaling}{b}. The state-of-the-art collection efficiency of 4.35\% has been experimentally achieved in the previous work \cite{crain2019high}, and with this value, the detection error can be significantly reduced to around $2\times10^{-4}$. By further including the information of photon arrival time and combining techniques such as machine learning \cite{PhysRevApplied.12.014038, seif2018machine, jeong2023using}, the detection error can be suppressed to below $10^{-4}$. 

\subsection{Feasibility for mid-circuit measurement}

\begin{figure}[htbp]
  \includegraphics[scale = 1.0]{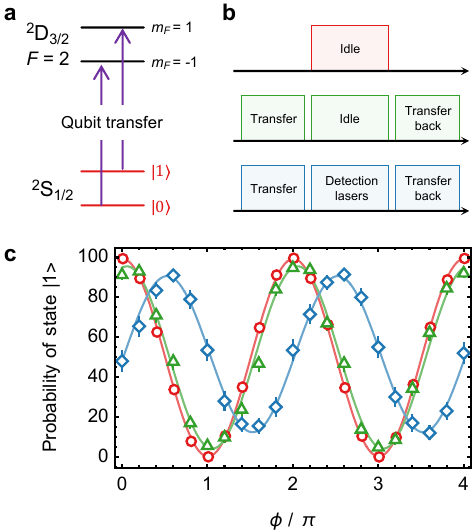}
  \caption{
    \label{fig:mid-circuit-measure}
    Scheme to realize the mid-circuit measurement.
    \textbf{a} Qubit state transfer. The qubit encoded in the ground-manifold hyperfine levels can be coherently transferred to a pair of Zeeman levels belong to $^2D_{3/2}$, or vice versa.
    \textbf{b} Experimental sequences to examine the coherence perservation during the mid-circuit measurement.
    \textbf{c} Results of qubit coherence. Each colored result corresponds to an experimental sequence of the same color in \textbf{b}. All the open markers represent experimental data points, while the solid lines depict corresponding fitting results. The contrast for each line is as follows, 99(1)\% (red), 91(3)\%(green), 78(2)\% (blue).
  }
\end{figure}

In addition to utilizing $^2D_{3/2}$ states for achieving high-fidelity shelving detection, there are straightforward extensions for realizing mid-circuit measurement. Specifically, we can protect the qubits that we do not want to measure by transferring their states, encoded in the hyperfine $^2S_{1/2}$ levels, to a pair of levels within $^2D_{3/2}\ket{F=2}$, as illustrated in \figref{fig:mid-circuit-measure}{a}. Consequently, these qubits remain concealed from all detection-related lasers and undisturbed.

We experimentally examine the above idea by investigating the coherence preservation of a single qubit in the superposition state when transferred to the $^2D_{3/2}$ levels and then measured. Initially, a single ion qubit is prepared in a superposition state of $(\ket{0} + \ket{1})/\sqrt{2}$. Subsequently, we independently test three following sequences (illustrated in \figref{fig:mid-circuit-measure}{b}):
(i) leaving the qubit idle for 70 $\mu$s.
(ii) transferring the qubit to the $^2D_{3/2}$ levels, waiting for 70 $\mu$s, and then transferring back.
(iii) transferring the qubit to the $^2D_{3/2}$ levels, applying detection-related lasers for 70 $\mu$s, and then transferring back. Finally, the remaining qubit coherence can be estimated by applying a phase-varied $\pi/2$-rotation and then extracting the oscillation amplitude. The results are depicted in \figref{fig:mid-circuit-measure}{c}.

In the reference scenario, the qubit, which consistently resides in the ground hyperfine levels, exhibits near-complete coherence preservation. For the case of the second sequence, the qubit coherence drops to 91(3)\%, due primarily to decoherence of the Zeeman levels induced by magnetic noise (around 700~$\mu$s coherence time). When the detection lasers are illuminated on the ion during its occupation of $^2D_{3/2}$ levels, the final coherence further decreases to 78(2)\%. We suspect the main error source to be the power and frequency fluctuations of the 935.2~nm laser, resulting in a noisy light shift affecting the Zeeman levels. Further improvements in the stability of both the magnetic field and the 935~nm laser could prove beneficial in enhancing the fidelity of the $^2D_{3/2}$-assisted mid-circuit measurement.

\section{
  \label{sec:conclusion}
  Conclusion
}
In summary, we have presented an experimenal demonstration of a high-fidelity detection scheme for $\yb$ qubit employing the $^2D_{3/2}$ manifold for shelving. Through effectively suppressing qubit state flips during detection, we have achieved a detection fidelity of 99.88(2)\% for individal $\yb$ qubit. Leveraging advancements in photon collection efficiency and implementing techniques such as machine learning, we anticipate pushing the detection fidelity for single ion qubit beyond 99.99\%. 

By further integrating low-cross-talk detection methods for multiple ion qubits, such as coupling ion-chain fluorescence to a fiber array, we can attain high-fidelity readout for multi-ion registers. This integration, combined with the feasibility of mid-circuit measurement via $^2D_{3/2}$ shelving, underscores the importance of our approach in advancing quantum error correction. Such capability of implementing mid-circuit measurements also hold promise for simulating open quantum systems and exploring nonlinear quantum dynamics \cite{PhysRevA.83.062317, PhysRevLett.127.020504, barreiro2011open, PhysRevLett.124.110503}. Additionally, we can also extend our approach to achieve high-fidelity $\yb$ qudits readout \cite{PhysRevA.107.052612}. 

Recent research has highlighted the potential of utilizing $^2D_{3/2}$ levels to achieve high-fidelity two-qubit gates and mitigate leakage errors in hyperfine $\yb$ qubits \cite{PhysRevA.103.012603, PhysRevLett.124.170501}. Combined with our scheme for improving ion qubit detection, we can fulfill multiple critical functions towards fault-tolerant quantum computation multiplexing the same narrow-linewidth transition driven by visible lasers. The capability of optical integration of such lasers on chip traps would further accelerate the scale-up of trapped-ion quantum processors \cite{PhysRevX.11.041033}.
\\

\begin{acknowledgments}
  This work was supported by the National Science Foundation of China (Grants No.~12004165), 
  the Shenzhen Science and Technology Program (Grants No.~RCYX20221008092901006), 
  the Guangdong Basic and Applied Basic Research Foundation (Grant No. 2022B1515120021), 
  the Guangdong Provincial Key Laboratory (Grant No.~2019B121203002).
\end{acknowledgments}


\section{appendix}

\subsection{Collection efficiency of imaging system}
Here we present a comprehensive analysis of the factors influencing photon collection efficiency in our experimental setup, detailed in \tabref{tab:collecting_efficiency}. We can find that the primary limiting factor is the N.A. of our objective lens. 

Theoretical estimation based on the static magnetic field strength utilized in our experiment suggests a saturated photon scattering rate of 23.6 photons/$\mu$s for the transition $^2S_{1/2}\ket{F = 1} \leftrightarrow {^2P_{1/2}\ket{F = 0}}$ \cite{PhysRevA.65.033413}. Our practical implementation yields a collection rate of 0.180(1) photons/$\mu$s, indicating an overall collection efficiency of 0.763(4)\%. Moreover, the background count rate of our setup is estimated to 6.3~/s, primarily contributed by the dark count of the PMT.

\begin{table}[htbp]
  \centering
  \tabcolsep = 0.2cm
  \renewcommand\arraystretch{1.5}
  \caption{\label{tab:collecting_efficiency}%
  Contribution for photon collection effeciency
  }
    \begin{tabular}{c|c}
      \hline\hline
       & Contribution \\
      \hline
      Solid angle of objective lens & $\sim$4\% \\
      Transmission of objective lens & $\sim$85\% \\
      Transmission of optical filter & $\sim$95\% \\
      Fiber coupling effeciency & $\sim$75\% \\
      Quantum efficiency of PMT & $\sim$32\% \\
      \hline
      Estimated total efficiency & 0.77\% \\
      \hline
      \hline
    \end{tabular}
\end{table}

\subsection{Theoretical derivation of leakage suppression}

In the main text, we simplify the leakage channel of the $\ket{1}$ state as $^2S_{1/2}\ket{F = 1} \rightarrow {^2P_{1/2}\ket{F = 1}} \rightarrow {^2D_{3/ 2}\ket{F = 2}}$. However, once the qubit in $^2S_{1/2}\ket{F = 1}$ levels is off-resonantly excited to $^2P_{1/2}\ket{F = 1}$, it has a high probability of decaying to $^2S_{1/2}\ket{F = 0}$. To pump out state occupied in this level, we utilize a resonant transition from $^2S_{1/2}\ket{F = 0}$ to $^2P_{1/2}\ket{F = 1}$, which introduces another chance for the qubit state to leak to $^2D_{3/2}\ket{F = 2}$. Therefore, we estimate the overall probability of the state in $^2S_{1/2}\ket{F = 1}$ leaking to $^2D_{3/2}\ket{F = 2}$ as follows:
\begin{eqnarray}
  p_\mathrm{shelving} = & &p_\mathrm{off-resonant}~p_{P \rightarrow D} \nonumber\\ 
  & + &p_\mathrm{off-resonant} ~ p_{P \rightarrow S} ~ p_{P \rightarrow D} \nonumber\\
  & + &p_\mathrm{off-resonant} ~ p_{P \rightarrow S}^2 ~ p_{P \rightarrow D} \nonumber\\
  & + & \dots \nonumber\\
  = & & p_\mathrm{off-resonant} \dfrac{p_{P \rightarrow D}}{1 - p_{P \rightarrow S}}.
\end{eqnarray}
Here, $p_{P \rightarrow S}$ represents the probability of the state in $^2P_{1/2}\ket{F = 1}$ decaying to $^2S_{1/2}\ket{F = 0}$, with a value of $0.995\times1/3$. $p_{P \rightarrow D}$ is the probability of the state in $^2P_{1/2}\ket{F = 1}$ decaying to $^2D_{3/2}\ket{F = 2}$, with a value of $0.005\times5/6$. The leakage suppression factor, defined as:
\begin{eqnarray}
  r & = & \dfrac{p_\mathrm{hyperfine}}{p_\mathrm{shelving}} \nonumber \\
  & = & \dfrac{p_\mathrm{off-resonant}~1/3}{p_\mathrm{off-resonant}~p_{P \rightarrow D} / (1 - p_{P \rightarrow S})} \nonumber \\
  & = & \dfrac{1 - p_{P \rightarrow S}}{3~p_{P \rightarrow D}}
\end{eqnarray}
is finally estimated to be 53.

\bibliography{Refs}

\begin{thebibliography}{40}%
\makeatletter
\providecommand \@ifxundefined [1]{%
 \@ifx{#1\undefined}
}%
\providecommand \@ifnum [1]{%
 \ifnum #1\expandafter \@firstoftwo
 \else \expandafter \@secondoftwo
 \fi
}%
\providecommand \@ifx [1]{%
 \ifx #1\expandafter \@firstoftwo
 \else \expandafter \@secondoftwo
 \fi
}%
\providecommand \natexlab [1]{#1}%
\providecommand \enquote  [1]{``#1''}%
\providecommand \bibnamefont  [1]{#1}%
\providecommand \bibfnamefont [1]{#1}%
\providecommand \citenamefont [1]{#1}%
\providecommand \href@noop [0]{\@secondoftwo}%
\providecommand \href [0]{\begingroup \@sanitize@url \@href}%
\providecommand \@href[1]{\@@startlink{#1}\@@href}%
\providecommand \@@href[1]{\endgroup#1\@@endlink}%
\providecommand \@sanitize@url [0]{\catcode `\\12\catcode `\$12\catcode
  `\&12\catcode `\#12\catcode `\^12\catcode `\_12\catcode `\%12\relax}%
\providecommand \@@startlink[1]{}%
\providecommand \@@endlink[0]{}%
\providecommand \url  [0]{\begingroup\@sanitize@url \@url }%
\providecommand \@url [1]{\endgroup\@href {#1}{\urlprefix }}%
\providecommand \urlprefix  [0]{URL }%
\providecommand \Eprint [0]{\href }%
\providecommand \doibase [0]{https://doi.org/}%
\providecommand \selectlanguage [0]{\@gobble}%
\providecommand \bibinfo  [0]{\@secondoftwo}%
\providecommand \bibfield  [0]{\@secondoftwo}%
\providecommand \translation [1]{[#1]}%
\providecommand \BibitemOpen [0]{}%
\providecommand \bibitemStop [0]{}%
\providecommand \bibitemNoStop [0]{.\EOS\space}%
\providecommand \EOS [0]{\spacefactor3000\relax}%
\providecommand \BibitemShut  [1]{\csname bibitem#1\endcsname}%
\let\auto@bib@innerbib\@empty
\bibitem [{\citenamefont {Kielpinski}\ \emph {et~al.}(2002)\citenamefont
  {Kielpinski}, \citenamefont {Monroe},\ and\ \citenamefont
  {Wineland}}]{kielpinski2002architecture}%
  \BibitemOpen
  \bibfield  {author} {\bibinfo {author} {\bibfnamefont {D.}~\bibnamefont
  {Kielpinski}}, \bibinfo {author} {\bibfnamefont {C.}~\bibnamefont {Monroe}},\
  and\ \bibinfo {author} {\bibfnamefont {D.~J.}\ \bibnamefont {Wineland}},\
  }\bibfield  {title} {\bibinfo {title} {Architecture for a large-scale
  ion-trap quantum computer},\ }\href {https://doi.org/10.1038/nature00784}
  {\bibfield  {journal} {\bibinfo  {journal} {Nature}\ }\textbf {\bibinfo
  {volume} {417}},\ \bibinfo {pages} {709} (\bibinfo {year}
  {2002})}\BibitemShut {NoStop}%
\bibitem [{\citenamefont {Monroe}\ and\ \citenamefont
  {Kim}(2013)}]{monroe2013scaling}%
  \BibitemOpen
  \bibfield  {author} {\bibinfo {author} {\bibfnamefont {C.}~\bibnamefont
  {Monroe}}\ and\ \bibinfo {author} {\bibfnamefont {J.}~\bibnamefont {Kim}},\
  }\bibfield  {title} {\bibinfo {title} {Scaling the ion trap quantum
  processor},\ }\href {https://doi.org/10.1126/science.1231298} {\bibfield
  {journal} {\bibinfo  {journal} {Science}\ }\textbf {\bibinfo {volume}
  {339}},\ \bibinfo {pages} {1164} (\bibinfo {year} {2013})}\BibitemShut
  {NoStop}%
\bibitem [{\citenamefont {Brown}\ \emph {et~al.}(2016)\citenamefont {Brown},
  \citenamefont {Kim},\ and\ \citenamefont {Monroe}}]{brown2016co}%
  \BibitemOpen
  \bibfield  {author} {\bibinfo {author} {\bibfnamefont {K.~R.}\ \bibnamefont
  {Brown}}, \bibinfo {author} {\bibfnamefont {J.}~\bibnamefont {Kim}},\ and\
  \bibinfo {author} {\bibfnamefont {C.}~\bibnamefont {Monroe}},\ }\bibfield
  {title} {\bibinfo {title} {Co-designing a scalable quantum computer with
  trapped atomic ions},\ }\href {https://doi.org/10.1038/npjqi.2016.34}
  {\bibfield  {journal} {\bibinfo  {journal} {npj Quantum Inf.}\ }\textbf
  {\bibinfo {volume} {2}},\ \bibinfo {pages} {1} (\bibinfo {year}
  {2016})}\BibitemShut {NoStop}%
\bibitem [{\citenamefont {Wang}\ \emph {et~al.}(2017)\citenamefont {Wang},
  \citenamefont {Um}, \citenamefont {Zhang}, \citenamefont {An}, \citenamefont
  {Lyu}, \citenamefont {Zhang}, \citenamefont {Duan}, \citenamefont {Yum},\
  and\ \citenamefont {Kim}}]{wang2017single}%
  \BibitemOpen
  \bibfield  {author} {\bibinfo {author} {\bibfnamefont {Y.}~\bibnamefont
  {Wang}}, \bibinfo {author} {\bibfnamefont {M.}~\bibnamefont {Um}}, \bibinfo
  {author} {\bibfnamefont {J.}~\bibnamefont {Zhang}}, \bibinfo {author}
  {\bibfnamefont {S.}~\bibnamefont {An}}, \bibinfo {author} {\bibfnamefont
  {M.}~\bibnamefont {Lyu}}, \bibinfo {author} {\bibfnamefont {J.-N.}\
  \bibnamefont {Zhang}}, \bibinfo {author} {\bibfnamefont {L.-M.}\ \bibnamefont
  {Duan}}, \bibinfo {author} {\bibfnamefont {D.}~\bibnamefont {Yum}},\ and\
  \bibinfo {author} {\bibfnamefont {K.}~\bibnamefont {Kim}},\ }\bibfield
  {title} {\bibinfo {title} {Single-qubit quantum memory exceeding ten-minute
  coherence time},\ }\href {https://doi.org/10.1038/s41566-017-0052-9}
  {\bibfield  {journal} {\bibinfo  {journal} {Nat. Photonics}\ }\textbf
  {\bibinfo {volume} {11}},\ \bibinfo {pages} {646} (\bibinfo {year}
  {2017})}\BibitemShut {NoStop}%
\bibitem [{\citenamefont {Wang}\ \emph {et~al.}(2021)\citenamefont {Wang},
  \citenamefont {Luan}, \citenamefont {Qiao}, \citenamefont {Um}, \citenamefont
  {Zhang}, \citenamefont {Wang}, \citenamefont {Yuan}, \citenamefont {Gu},
  \citenamefont {Zhang},\ and\ \citenamefont {Kim}}]{wang2021single}%
  \BibitemOpen
  \bibfield  {author} {\bibinfo {author} {\bibfnamefont {P.}~\bibnamefont
  {Wang}}, \bibinfo {author} {\bibfnamefont {C.-Y.}\ \bibnamefont {Luan}},
  \bibinfo {author} {\bibfnamefont {M.}~\bibnamefont {Qiao}}, \bibinfo {author}
  {\bibfnamefont {M.}~\bibnamefont {Um}}, \bibinfo {author} {\bibfnamefont
  {J.}~\bibnamefont {Zhang}}, \bibinfo {author} {\bibfnamefont
  {Y.}~\bibnamefont {Wang}}, \bibinfo {author} {\bibfnamefont {X.}~\bibnamefont
  {Yuan}}, \bibinfo {author} {\bibfnamefont {M.}~\bibnamefont {Gu}}, \bibinfo
  {author} {\bibfnamefont {J.}~\bibnamefont {Zhang}},\ and\ \bibinfo {author}
  {\bibfnamefont {K.}~\bibnamefont {Kim}},\ }\bibfield  {title} {\bibinfo
  {title} {Single ion qubit with estimated coherence time exceeding one hour},\
  }\href {https://doi.org/10.1038/s41467-020-20330-w} {\bibfield  {journal}
  {\bibinfo  {journal} {Nat. Commun.}\ }\textbf {\bibinfo {volume} {12}},\
  \bibinfo {pages} {233} (\bibinfo {year} {2021})}\BibitemShut {NoStop}%
\bibitem [{\citenamefont {Leu}\ \emph {et~al.}(2023)\citenamefont {Leu},
  \citenamefont {Gely}, \citenamefont {Weber}, \citenamefont {Smith},
  \citenamefont {Nadlinger},\ and\ \citenamefont
  {Lucas}}]{PhysRevLett.131.120601}%
  \BibitemOpen
  \bibfield  {author} {\bibinfo {author} {\bibfnamefont {A.~D.}\ \bibnamefont
  {Leu}}, \bibinfo {author} {\bibfnamefont {M.~F.}\ \bibnamefont {Gely}},
  \bibinfo {author} {\bibfnamefont {M.~A.}\ \bibnamefont {Weber}}, \bibinfo
  {author} {\bibfnamefont {M.~C.}\ \bibnamefont {Smith}}, \bibinfo {author}
  {\bibfnamefont {D.~P.}\ \bibnamefont {Nadlinger}},\ and\ \bibinfo {author}
  {\bibfnamefont {D.~M.}\ \bibnamefont {Lucas}},\ }\bibfield  {title} {\bibinfo
  {title} {{Fast, High-Fidelity Addressed Single-Qubit Gates Using Efficient
  Composite Pulse Sequences}},\ }\href
  {https://doi.org/10.1103/PhysRevLett.131.120601} {\bibfield  {journal}
  {\bibinfo  {journal} {Phys. Rev. Lett.}\ }\textbf {\bibinfo {volume} {131}},\
  \bibinfo {pages} {120601} (\bibinfo {year} {2023})}\BibitemShut {NoStop}%
\bibitem [{\citenamefont {Ballance}\ \emph {et~al.}(2016)\citenamefont
  {Ballance}, \citenamefont {Harty}, \citenamefont {Linke}, \citenamefont
  {Sepiol},\ and\ \citenamefont {Lucas}}]{PhysRevLett.117.060504}%
  \BibitemOpen
  \bibfield  {author} {\bibinfo {author} {\bibfnamefont {C.~J.}\ \bibnamefont
  {Ballance}}, \bibinfo {author} {\bibfnamefont {T.~P.}\ \bibnamefont {Harty}},
  \bibinfo {author} {\bibfnamefont {N.~M.}\ \bibnamefont {Linke}}, \bibinfo
  {author} {\bibfnamefont {M.~A.}\ \bibnamefont {Sepiol}},\ and\ \bibinfo
  {author} {\bibfnamefont {D.~M.}\ \bibnamefont {Lucas}},\ }\bibfield  {title}
  {\bibinfo {title} {{High-Fidelity Quantum Logic Gates Using Trapped-Ion
  Hyperfine Qubits}},\ }\href {https://doi.org/10.1103/PhysRevLett.117.060504}
  {\bibfield  {journal} {\bibinfo  {journal} {Phys. Rev. Lett.}\ }\textbf
  {\bibinfo {volume} {117}},\ \bibinfo {pages} {060504} (\bibinfo {year}
  {2016})}\BibitemShut {NoStop}%
\bibitem [{\citenamefont {Gaebler}\ \emph {et~al.}(2016)\citenamefont
  {Gaebler}, \citenamefont {Tan}, \citenamefont {Lin}, \citenamefont {Wan},
  \citenamefont {Bowler}, \citenamefont {Keith}, \citenamefont {Glancy},
  \citenamefont {Coakley}, \citenamefont {Knill}, \citenamefont {Leibfried},\
  and\ \citenamefont {Wineland}}]{PhysRevLett.117.060505}%
  \BibitemOpen
  \bibfield  {author} {\bibinfo {author} {\bibfnamefont {J.~P.}\ \bibnamefont
  {Gaebler}}, \bibinfo {author} {\bibfnamefont {T.~R.}\ \bibnamefont {Tan}},
  \bibinfo {author} {\bibfnamefont {Y.}~\bibnamefont {Lin}}, \bibinfo {author}
  {\bibfnamefont {Y.}~\bibnamefont {Wan}}, \bibinfo {author} {\bibfnamefont
  {R.}~\bibnamefont {Bowler}}, \bibinfo {author} {\bibfnamefont {A.~C.}\
  \bibnamefont {Keith}}, \bibinfo {author} {\bibfnamefont {S.}~\bibnamefont
  {Glancy}}, \bibinfo {author} {\bibfnamefont {K.}~\bibnamefont {Coakley}},
  \bibinfo {author} {\bibfnamefont {E.}~\bibnamefont {Knill}}, \bibinfo
  {author} {\bibfnamefont {D.}~\bibnamefont {Leibfried}},\ and\ \bibinfo
  {author} {\bibfnamefont {D.~J.}\ \bibnamefont {Wineland}},\ }\bibfield
  {title} {\bibinfo {title} {{High-Fidelity Universal Gate Set for
  ${^{9}\mathrm{Be}}^{+}$ Ion Qubits}},\ }\href
  {https://doi.org/10.1103/PhysRevLett.117.060505} {\bibfield  {journal}
  {\bibinfo  {journal} {Phys. Rev. Lett.}\ }\textbf {\bibinfo {volume} {117}},\
  \bibinfo {pages} {060505} (\bibinfo {year} {2016})}\BibitemShut {NoStop}%
\bibitem [{\citenamefont {Clark}\ \emph {et~al.}(2021)\citenamefont {Clark},
  \citenamefont {Tinkey}, \citenamefont {Sawyer}, \citenamefont {Meier},
  \citenamefont {Burkhardt}, \citenamefont {Seck}, \citenamefont {Shappert},
  \citenamefont {Guise}, \citenamefont {Volin}, \citenamefont {Fallek},
  \citenamefont {Hayden}, \citenamefont {Rellergert},\ and\ \citenamefont
  {Brown}}]{PhysRevLett.127.130505}%
  \BibitemOpen
  \bibfield  {author} {\bibinfo {author} {\bibfnamefont {C.~R.}\ \bibnamefont
  {Clark}}, \bibinfo {author} {\bibfnamefont {H.~N.}\ \bibnamefont {Tinkey}},
  \bibinfo {author} {\bibfnamefont {B.~C.}\ \bibnamefont {Sawyer}}, \bibinfo
  {author} {\bibfnamefont {A.~M.}\ \bibnamefont {Meier}}, \bibinfo {author}
  {\bibfnamefont {K.~A.}\ \bibnamefont {Burkhardt}}, \bibinfo {author}
  {\bibfnamefont {C.~M.}\ \bibnamefont {Seck}}, \bibinfo {author}
  {\bibfnamefont {C.~M.}\ \bibnamefont {Shappert}}, \bibinfo {author}
  {\bibfnamefont {N.~D.}\ \bibnamefont {Guise}}, \bibinfo {author}
  {\bibfnamefont {C.~E.}\ \bibnamefont {Volin}}, \bibinfo {author}
  {\bibfnamefont {S.~D.}\ \bibnamefont {Fallek}}, \bibinfo {author}
  {\bibfnamefont {H.~T.}\ \bibnamefont {Hayden}}, \bibinfo {author}
  {\bibfnamefont {W.~G.}\ \bibnamefont {Rellergert}},\ and\ \bibinfo {author}
  {\bibfnamefont {K.~R.}\ \bibnamefont {Brown}},\ }\bibfield  {title} {\bibinfo
  {title} {{High-Fidelity Bell-State Preparation with $^{40}{\mathrm{Ca}}^{+}$
  Optical Qubits}},\ }\href {https://doi.org/10.1103/PhysRevLett.127.130505}
  {\bibfield  {journal} {\bibinfo  {journal} {Phys. Rev. Lett.}\ }\textbf
  {\bibinfo {volume} {127}},\ \bibinfo {pages} {130505} (\bibinfo {year}
  {2021})}\BibitemShut {NoStop}%
\bibitem [{\citenamefont {Srinivas}\ \emph {et~al.}(2021)\citenamefont
  {Srinivas}, \citenamefont {Burd}, \citenamefont {Knaack}, \citenamefont
  {Sutherland}, \citenamefont {Kwiatkowski}, \citenamefont {Glancy},
  \citenamefont {Knill}, \citenamefont {Wineland}, \citenamefont {Leibfried},
  \citenamefont {Wilson} \emph {et~al.}}]{srinivas2021high}%
  \BibitemOpen
  \bibfield  {author} {\bibinfo {author} {\bibfnamefont {R.}~\bibnamefont
  {Srinivas}}, \bibinfo {author} {\bibfnamefont {S.}~\bibnamefont {Burd}},
  \bibinfo {author} {\bibfnamefont {H.}~\bibnamefont {Knaack}}, \bibinfo
  {author} {\bibfnamefont {R.}~\bibnamefont {Sutherland}}, \bibinfo {author}
  {\bibfnamefont {A.}~\bibnamefont {Kwiatkowski}}, \bibinfo {author}
  {\bibfnamefont {S.}~\bibnamefont {Glancy}}, \bibinfo {author} {\bibfnamefont
  {E.}~\bibnamefont {Knill}}, \bibinfo {author} {\bibfnamefont
  {D.}~\bibnamefont {Wineland}}, \bibinfo {author} {\bibfnamefont
  {D.}~\bibnamefont {Leibfried}}, \bibinfo {author} {\bibfnamefont {A.~C.}\
  \bibnamefont {Wilson}}, \emph {et~al.},\ }\bibfield  {title} {\bibinfo
  {title} {{High-fidelity laser-free universal control of trapped ion
  qubits}},\ }\href {https://doi.org/10.1038/s41586-021-03809-4} {\bibfield
  {journal} {\bibinfo  {journal} {Nature}\ }\textbf {\bibinfo {volume} {597}},\
  \bibinfo {pages} {209} (\bibinfo {year} {2021})}\BibitemShut {NoStop}%
\bibitem [{\citenamefont {Lu}\ \emph {et~al.}(2019)\citenamefont {Lu},
  \citenamefont {Zhang}, \citenamefont {Zhang}, \citenamefont {Chen},
  \citenamefont {Shen}, \citenamefont {Zhang}, \citenamefont {Zhang},\ and\
  \citenamefont {Kim}}]{lu2019global}%
  \BibitemOpen
  \bibfield  {author} {\bibinfo {author} {\bibfnamefont {Y.}~\bibnamefont
  {Lu}}, \bibinfo {author} {\bibfnamefont {S.}~\bibnamefont {Zhang}}, \bibinfo
  {author} {\bibfnamefont {K.}~\bibnamefont {Zhang}}, \bibinfo {author}
  {\bibfnamefont {W.}~\bibnamefont {Chen}}, \bibinfo {author} {\bibfnamefont
  {Y.}~\bibnamefont {Shen}}, \bibinfo {author} {\bibfnamefont {J.}~\bibnamefont
  {Zhang}}, \bibinfo {author} {\bibfnamefont {J.-N.}\ \bibnamefont {Zhang}},\
  and\ \bibinfo {author} {\bibfnamefont {K.}~\bibnamefont {Kim}},\ }\bibfield
  {title} {\bibinfo {title} {Global entangling gates on arbitrary ion qubits},\
  }\href {https://doi.org/10.1038/s41586-019-1428-4} {\bibfield  {journal}
  {\bibinfo  {journal} {Nature}\ }\textbf {\bibinfo {volume} {572}},\ \bibinfo
  {pages} {363} (\bibinfo {year} {2019})}\BibitemShut {NoStop}%
\bibitem [{\citenamefont {Figgatt}\ \emph {et~al.}(2019)\citenamefont
  {Figgatt}, \citenamefont {Ostrander}, \citenamefont {Linke}, \citenamefont
  {Landsman}, \citenamefont {Zhu}, \citenamefont {Maslov},\ and\ \citenamefont
  {Monroe}}]{figgatt2019parallel}%
  \BibitemOpen
  \bibfield  {author} {\bibinfo {author} {\bibfnamefont {C.}~\bibnamefont
  {Figgatt}}, \bibinfo {author} {\bibfnamefont {A.}~\bibnamefont {Ostrander}},
  \bibinfo {author} {\bibfnamefont {N.~M.}\ \bibnamefont {Linke}}, \bibinfo
  {author} {\bibfnamefont {K.~A.}\ \bibnamefont {Landsman}}, \bibinfo {author}
  {\bibfnamefont {D.}~\bibnamefont {Zhu}}, \bibinfo {author} {\bibfnamefont
  {D.}~\bibnamefont {Maslov}},\ and\ \bibinfo {author} {\bibfnamefont
  {C.}~\bibnamefont {Monroe}},\ }\bibfield  {title} {\bibinfo {title} {Parallel
  entangling operations on a universal ion-trap quantum computer},\ }\href
  {https://doi.org/10.1038/s41586-019-1427-5} {\bibfield  {journal} {\bibinfo
  {journal} {Nature}\ }\textbf {\bibinfo {volume} {572}},\ \bibinfo {pages}
  {368} (\bibinfo {year} {2019})}\BibitemShut {NoStop}%
\bibitem [{\citenamefont {DiVincenzo}(2000)}]{divincenzo2000physical}%
  \BibitemOpen
  \bibfield  {author} {\bibinfo {author} {\bibfnamefont {D.~P.}\ \bibnamefont
  {DiVincenzo}},\ }\bibfield  {title} {\bibinfo {title} {The physical
  implementation of quantum computation},\ }\href
  {https://doi.org/10.1002/1521-3978%28200009%2948%3A9/11<771%3A%3AAID-PROP771>3.0.CO%3B2-E}
  {\bibfield  {journal} {\bibinfo  {journal} {Fortschr. Phys.}\ }\textbf
  {\bibinfo {volume} {48}},\ \bibinfo {pages} {771} (\bibinfo {year}
  {2000})}\BibitemShut {NoStop}%
\bibitem [{\citenamefont {Egan}\ \emph {et~al.}(2021)\citenamefont {Egan},
  \citenamefont {Debroy}, \citenamefont {Noel}, \citenamefont {Risinger},
  \citenamefont {Zhu}, \citenamefont {Biswas}, \citenamefont {Newman},
  \citenamefont {Li}, \citenamefont {Brown}, \citenamefont {Cetina} \emph
  {et~al.}}]{egan2021fault}%
  \BibitemOpen
  \bibfield  {author} {\bibinfo {author} {\bibfnamefont {L.}~\bibnamefont
  {Egan}}, \bibinfo {author} {\bibfnamefont {D.~M.}\ \bibnamefont {Debroy}},
  \bibinfo {author} {\bibfnamefont {C.}~\bibnamefont {Noel}}, \bibinfo {author}
  {\bibfnamefont {A.}~\bibnamefont {Risinger}}, \bibinfo {author}
  {\bibfnamefont {D.}~\bibnamefont {Zhu}}, \bibinfo {author} {\bibfnamefont
  {D.}~\bibnamefont {Biswas}}, \bibinfo {author} {\bibfnamefont
  {M.}~\bibnamefont {Newman}}, \bibinfo {author} {\bibfnamefont
  {M.}~\bibnamefont {Li}}, \bibinfo {author} {\bibfnamefont {K.~R.}\
  \bibnamefont {Brown}}, \bibinfo {author} {\bibfnamefont {M.}~\bibnamefont
  {Cetina}}, \emph {et~al.},\ }\bibfield  {title} {\bibinfo {title}
  {Fault-tolerant control of an error-corrected qubit},\ }\href
  {https://doi.org/10.1038/s41586-021-03928-y} {\bibfield  {journal} {\bibinfo
  {journal} {Nature}\ }\textbf {\bibinfo {volume} {598}},\ \bibinfo {pages}
  {281} (\bibinfo {year} {2021})}\BibitemShut {NoStop}%
\bibitem [{\citenamefont {Postler}\ \emph {et~al.}(2022)\citenamefont
  {Postler}, \citenamefont {Heu$\beta$en}, \citenamefont {Pogorelov},
  \citenamefont {Rispler}, \citenamefont {Feldker}, \citenamefont {Meth},
  \citenamefont {Marciniak}, \citenamefont {Stricker}, \citenamefont
  {Ringbauer}, \citenamefont {Blatt} \emph
  {et~al.}}]{postler2022demonstration}%
  \BibitemOpen
  \bibfield  {author} {\bibinfo {author} {\bibfnamefont {L.}~\bibnamefont
  {Postler}}, \bibinfo {author} {\bibfnamefont {S.}~\bibnamefont
  {Heu$\beta$en}}, \bibinfo {author} {\bibfnamefont {I.}~\bibnamefont
  {Pogorelov}}, \bibinfo {author} {\bibfnamefont {M.}~\bibnamefont {Rispler}},
  \bibinfo {author} {\bibfnamefont {T.}~\bibnamefont {Feldker}}, \bibinfo
  {author} {\bibfnamefont {M.}~\bibnamefont {Meth}}, \bibinfo {author}
  {\bibfnamefont {C.~D.}\ \bibnamefont {Marciniak}}, \bibinfo {author}
  {\bibfnamefont {R.}~\bibnamefont {Stricker}}, \bibinfo {author}
  {\bibfnamefont {M.}~\bibnamefont {Ringbauer}}, \bibinfo {author}
  {\bibfnamefont {R.}~\bibnamefont {Blatt}}, \emph {et~al.},\ }\bibfield
  {title} {\bibinfo {title} {Demonstration of fault-tolerant universal quantum
  gate operations},\ }\href {https://doi.org/10.1038/s41586-022-04721-1}
  {\bibfield  {journal} {\bibinfo  {journal} {Nature}\ }\textbf {\bibinfo
  {volume} {605}},\ \bibinfo {pages} {675} (\bibinfo {year}
  {2022})}\BibitemShut {NoStop}%
\bibitem [{\citenamefont {DeCross}\ \emph {et~al.}(2023)\citenamefont
  {DeCross}, \citenamefont {Chertkov}, \citenamefont {Kohagen},\ and\
  \citenamefont {Foss-Feig}}]{PhysRevX.13.041057}%
  \BibitemOpen
  \bibfield  {author} {\bibinfo {author} {\bibfnamefont {M.}~\bibnamefont
  {DeCross}}, \bibinfo {author} {\bibfnamefont {E.}~\bibnamefont {Chertkov}},
  \bibinfo {author} {\bibfnamefont {M.}~\bibnamefont {Kohagen}},\ and\ \bibinfo
  {author} {\bibfnamefont {M.}~\bibnamefont {Foss-Feig}},\ }\bibfield  {title}
  {\bibinfo {title} {{Qubit-Reuse Compilation with Mid-Circuit Measurement and
  Reset}},\ }\href {https://doi.org/10.1103/PhysRevX.13.041057} {\bibfield
  {journal} {\bibinfo  {journal} {Phys. Rev. X}\ }\textbf {\bibinfo {volume}
  {13}},\ \bibinfo {pages} {041057} (\bibinfo {year} {2023})}\BibitemShut
  {NoStop}%
\bibitem [{\citenamefont {Wineland}\ \emph {et~al.}(1995)\citenamefont
  {Wineland}, \citenamefont {Bergquist}, \citenamefont {Bollinger},\ and\
  \citenamefont {Itano}}]{wineland1995quantum}%
  \BibitemOpen
  \bibfield  {author} {\bibinfo {author} {\bibfnamefont {D.~J.}\ \bibnamefont
  {Wineland}}, \bibinfo {author} {\bibfnamefont {J.~C.}\ \bibnamefont
  {Bergquist}}, \bibinfo {author} {\bibfnamefont {J.~J.}\ \bibnamefont
  {Bollinger}},\ and\ \bibinfo {author} {\bibfnamefont {W.~M.}\ \bibnamefont
  {Itano}},\ }\bibfield  {title} {\bibinfo {title} {Quantum effects in
  measurements on trapped ions},\ }\href
  {https://doi.org/10.1088/0031-8949/1995/T59/039} {\bibfield  {journal}
  {\bibinfo  {journal} {Phys. Scr.}\ }\textbf {\bibinfo {volume} {1995}},\
  \bibinfo {pages} {286} (\bibinfo {year} {1995})}\BibitemShut {NoStop}%
\bibitem [{\citenamefont {Olmschenk}\ \emph {et~al.}(2007)\citenamefont
  {Olmschenk}, \citenamefont {Younge}, \citenamefont {Moehring}, \citenamefont
  {Matsukevich}, \citenamefont {Maunz},\ and\ \citenamefont
  {Monroe}}]{PhysRevA.76.052314}%
  \BibitemOpen
  \bibfield  {author} {\bibinfo {author} {\bibfnamefont {S.}~\bibnamefont
  {Olmschenk}}, \bibinfo {author} {\bibfnamefont {K.~C.}\ \bibnamefont
  {Younge}}, \bibinfo {author} {\bibfnamefont {D.~L.}\ \bibnamefont
  {Moehring}}, \bibinfo {author} {\bibfnamefont {D.~N.}\ \bibnamefont
  {Matsukevich}}, \bibinfo {author} {\bibfnamefont {P.}~\bibnamefont {Maunz}},\
  and\ \bibinfo {author} {\bibfnamefont {C.}~\bibnamefont {Monroe}},\
  }\bibfield  {title} {\bibinfo {title} {{Manipulation and detection of a
  trapped ${\mathrm{Yb}}^{+}$ hyperfine qubit}},\ }\href
  {https://doi.org/10.1103/PhysRevA.76.052314} {\bibfield  {journal} {\bibinfo
  {journal} {Phys. Rev. A}\ }\textbf {\bibinfo {volume} {76}},\ \bibinfo
  {pages} {052314} (\bibinfo {year} {2007})}\BibitemShut {NoStop}%
\bibitem [{\citenamefont {Ejtemaee}\ \emph {et~al.}(2010)\citenamefont
  {Ejtemaee}, \citenamefont {Thomas},\ and\ \citenamefont
  {Haljan}}]{PhysRevA.82.063419}%
  \BibitemOpen
  \bibfield  {author} {\bibinfo {author} {\bibfnamefont {S.}~\bibnamefont
  {Ejtemaee}}, \bibinfo {author} {\bibfnamefont {R.}~\bibnamefont {Thomas}},\
  and\ \bibinfo {author} {\bibfnamefont {P.~C.}\ \bibnamefont {Haljan}},\
  }\bibfield  {title} {\bibinfo {title} {{Optimization of Yb${}^{+}$
  fluorescence and hyperfine-qubit detection}},\ }\href
  {https://doi.org/10.1103/PhysRevA.82.063419} {\bibfield  {journal} {\bibinfo
  {journal} {Phys. Rev. A}\ }\textbf {\bibinfo {volume} {82}},\ \bibinfo
  {pages} {063419} (\bibinfo {year} {2010})}\BibitemShut {NoStop}%
\bibitem [{\citenamefont {Noek}\ \emph {et~al.}(2013)\citenamefont {Noek},
  \citenamefont {Vrijsen}, \citenamefont {Gaultney}, \citenamefont {Mount},
  \citenamefont {Kim}, \citenamefont {Maunz},\ and\ \citenamefont
  {Kim}}]{noek2013high}%
  \BibitemOpen
  \bibfield  {author} {\bibinfo {author} {\bibfnamefont {R.}~\bibnamefont
  {Noek}}, \bibinfo {author} {\bibfnamefont {G.}~\bibnamefont {Vrijsen}},
  \bibinfo {author} {\bibfnamefont {D.}~\bibnamefont {Gaultney}}, \bibinfo
  {author} {\bibfnamefont {E.}~\bibnamefont {Mount}}, \bibinfo {author}
  {\bibfnamefont {T.}~\bibnamefont {Kim}}, \bibinfo {author} {\bibfnamefont
  {P.}~\bibnamefont {Maunz}},\ and\ \bibinfo {author} {\bibfnamefont
  {J.}~\bibnamefont {Kim}},\ }\bibfield  {title} {\bibinfo {title} {{High
  speed, high fidelity detection of an atomic hyperfine qubit}},\ }\href
  {https://doi.org/10.1364/OL.38.004735} {\bibfield  {journal} {\bibinfo
  {journal} {Opt. Lett.}\ }\textbf {\bibinfo {volume} {38}},\ \bibinfo {pages}
  {4735} (\bibinfo {year} {2013})}\BibitemShut {NoStop}%
\bibitem [{\citenamefont {Crain}\ \emph {et~al.}(2019)\citenamefont {Crain},
  \citenamefont {Cahall}, \citenamefont {Vrijsen}, \citenamefont {Wollman},
  \citenamefont {Shaw}, \citenamefont {Verma}, \citenamefont {Nam},\ and\
  \citenamefont {Kim}}]{crain2019high}%
  \BibitemOpen
  \bibfield  {author} {\bibinfo {author} {\bibfnamefont {S.}~\bibnamefont
  {Crain}}, \bibinfo {author} {\bibfnamefont {C.}~\bibnamefont {Cahall}},
  \bibinfo {author} {\bibfnamefont {G.}~\bibnamefont {Vrijsen}}, \bibinfo
  {author} {\bibfnamefont {E.~E.}\ \bibnamefont {Wollman}}, \bibinfo {author}
  {\bibfnamefont {M.~D.}\ \bibnamefont {Shaw}}, \bibinfo {author}
  {\bibfnamefont {V.~B.}\ \bibnamefont {Verma}}, \bibinfo {author}
  {\bibfnamefont {S.~W.}\ \bibnamefont {Nam}},\ and\ \bibinfo {author}
  {\bibfnamefont {J.}~\bibnamefont {Kim}},\ }\bibfield  {title} {\bibinfo
  {title} {{High-speed low-crosstalk detection of a ${}^{171}{\mathrm{Yb}}^{+}$
  qubit using superconducting nanowire single photon detectors}},\ }\href
  {https://doi.org/10.1038/s42005-019-0195-8} {\bibfield  {journal} {\bibinfo
  {journal} {Commun. Phys.}\ }\textbf {\bibinfo {volume} {2}},\ \bibinfo
  {pages} {97} (\bibinfo {year} {2019})}\BibitemShut {NoStop}%
\bibitem [{\citenamefont {Harty}\ \emph {et~al.}(2014)\citenamefont {Harty},
  \citenamefont {Allcock}, \citenamefont {Ballance}, \citenamefont {Guidoni},
  \citenamefont {Janacek}, \citenamefont {Linke}, \citenamefont {Stacey},\ and\
  \citenamefont {Lucas}}]{PhysRevLett.113.220501}%
  \BibitemOpen
  \bibfield  {author} {\bibinfo {author} {\bibfnamefont {T.~P.}\ \bibnamefont
  {Harty}}, \bibinfo {author} {\bibfnamefont {D.~T.~C.}\ \bibnamefont
  {Allcock}}, \bibinfo {author} {\bibfnamefont {C.~J.}\ \bibnamefont
  {Ballance}}, \bibinfo {author} {\bibfnamefont {L.}~\bibnamefont {Guidoni}},
  \bibinfo {author} {\bibfnamefont {H.~A.}\ \bibnamefont {Janacek}}, \bibinfo
  {author} {\bibfnamefont {N.~M.}\ \bibnamefont {Linke}}, \bibinfo {author}
  {\bibfnamefont {D.~N.}\ \bibnamefont {Stacey}},\ and\ \bibinfo {author}
  {\bibfnamefont {D.~M.}\ \bibnamefont {Lucas}},\ }\bibfield  {title} {\bibinfo
  {title} {{High-Fidelity Preparation, Gates, Memory, and Readout of a
  Trapped-Ion Quantum Bit}},\ }\href
  {https://doi.org/10.1103/PhysRevLett.113.220501} {\bibfield  {journal}
  {\bibinfo  {journal} {Phys. Rev. Lett.}\ }\textbf {\bibinfo {volume} {113}},\
  \bibinfo {pages} {220501} (\bibinfo {year} {2014})}\BibitemShut {NoStop}%
\bibitem [{\citenamefont {Keselman}\ \emph {et~al.}(2011)\citenamefont
  {Keselman}, \citenamefont {Glickman}, \citenamefont {Akerman}, \citenamefont
  {Kotler},\ and\ \citenamefont {Ozeri}}]{keselman2011high}%
  \BibitemOpen
  \bibfield  {author} {\bibinfo {author} {\bibfnamefont {A.}~\bibnamefont
  {Keselman}}, \bibinfo {author} {\bibfnamefont {Y.}~\bibnamefont {Glickman}},
  \bibinfo {author} {\bibfnamefont {N.}~\bibnamefont {Akerman}}, \bibinfo
  {author} {\bibfnamefont {S.}~\bibnamefont {Kotler}},\ and\ \bibinfo {author}
  {\bibfnamefont {R.}~\bibnamefont {Ozeri}},\ }\bibfield  {title} {\bibinfo
  {title} {{High-fidelity state detection and tomography of a single-ion Zeeman
  qubit}},\ }\href {https://doi.org/10.1088/1367-2630/13/7/073027} {\bibfield
  {journal} {\bibinfo  {journal} {New J. Phys.}\ }\textbf {\bibinfo {volume}
  {13}},\ \bibinfo {pages} {073027} (\bibinfo {year} {2011})}\BibitemShut
  {NoStop}%
\bibitem [{\citenamefont {Edmunds}\ \emph {et~al.}(2021)\citenamefont
  {Edmunds}, \citenamefont {Tan}, \citenamefont {Milne}, \citenamefont {Singh},
  \citenamefont {Biercuk},\ and\ \citenamefont {Hempel}}]{PhysRevA.104.012606}%
  \BibitemOpen
  \bibfield  {author} {\bibinfo {author} {\bibfnamefont {C.~L.}\ \bibnamefont
  {Edmunds}}, \bibinfo {author} {\bibfnamefont {T.~R.}\ \bibnamefont {Tan}},
  \bibinfo {author} {\bibfnamefont {A.~R.}\ \bibnamefont {Milne}}, \bibinfo
  {author} {\bibfnamefont {A.}~\bibnamefont {Singh}}, \bibinfo {author}
  {\bibfnamefont {M.~J.}\ \bibnamefont {Biercuk}},\ and\ \bibinfo {author}
  {\bibfnamefont {C.}~\bibnamefont {Hempel}},\ }\bibfield  {title} {\bibinfo
  {title} {{Scalable hyperfine qubit state detection via electron shelving in
  the ${}^{2}{D}_{5/2}$ and ${}^{2}{F}_{7/2}$ manifolds in
  ${}^{171}{\mathrm{Yb}}^{+}$}},\ }\href
  {https://doi.org/10.1103/PhysRevA.104.012606} {\bibfield  {journal} {\bibinfo
   {journal} {Phys. Rev. A}\ }\textbf {\bibinfo {volume} {104}},\ \bibinfo
  {pages} {012606} (\bibinfo {year} {2021})}\BibitemShut {NoStop}%
\bibitem [{\citenamefont {Ding}\ \emph {et~al.}(2019)\citenamefont {Ding},
  \citenamefont {Cui}, \citenamefont {Huang}, \citenamefont {Li}, \citenamefont
  {Tu},\ and\ \citenamefont {Guo}}]{PhysRevApplied.12.014038}%
  \BibitemOpen
  \bibfield  {author} {\bibinfo {author} {\bibfnamefont {Z.-H.}\ \bibnamefont
  {Ding}}, \bibinfo {author} {\bibfnamefont {J.-M.}\ \bibnamefont {Cui}},
  \bibinfo {author} {\bibfnamefont {Y.-F.}\ \bibnamefont {Huang}}, \bibinfo
  {author} {\bibfnamefont {C.-F.}\ \bibnamefont {Li}}, \bibinfo {author}
  {\bibfnamefont {T.}~\bibnamefont {Tu}},\ and\ \bibinfo {author}
  {\bibfnamefont {G.-C.}\ \bibnamefont {Guo}},\ }\bibfield  {title} {\bibinfo
  {title} {{Fast High-Fidelity Readout of a Single Trapped-Ion Qubit via
  Machine-Learning Methods}},\ }\href
  {https://doi.org/10.1103/PhysRevApplied.12.014038} {\bibfield  {journal}
  {\bibinfo  {journal} {Phys. Rev. Appl.}\ }\textbf {\bibinfo {volume} {12}},\
  \bibinfo {pages} {014038} (\bibinfo {year} {2019})}\BibitemShut {NoStop}%
\bibitem [{\citenamefont {Seif}\ \emph {et~al.}(2018)\citenamefont {Seif},
  \citenamefont {Landsman}, \citenamefont {Linke}, \citenamefont {Figgatt},
  \citenamefont {Monroe},\ and\ \citenamefont {Hafezi}}]{seif2018machine}%
  \BibitemOpen
  \bibfield  {author} {\bibinfo {author} {\bibfnamefont {A.}~\bibnamefont
  {Seif}}, \bibinfo {author} {\bibfnamefont {K.~A.}\ \bibnamefont {Landsman}},
  \bibinfo {author} {\bibfnamefont {N.~M.}\ \bibnamefont {Linke}}, \bibinfo
  {author} {\bibfnamefont {C.}~\bibnamefont {Figgatt}}, \bibinfo {author}
  {\bibfnamefont {C.}~\bibnamefont {Monroe}},\ and\ \bibinfo {author}
  {\bibfnamefont {M.}~\bibnamefont {Hafezi}},\ }\bibfield  {title} {\bibinfo
  {title} {Machine learning assisted readout of trapped-ion qubits},\ }\href
  {https://doi.org/10.1088/1361-6455/aad62b} {\bibfield  {journal} {\bibinfo
  {journal} {J. Phys. B At. Mol. Opt. Phys.}\ }\textbf {\bibinfo {volume}
  {51}},\ \bibinfo {pages} {174006} (\bibinfo {year} {2018})}\BibitemShut
  {NoStop}%
\bibitem [{\citenamefont {Jeong}\ \emph {et~al.}(2023)\citenamefont {Jeong},
  \citenamefont {Jung}, \citenamefont {Kim} \emph {et~al.}}]{jeong2023using}%
  \BibitemOpen
  \bibfield  {author} {\bibinfo {author} {\bibfnamefont {J.}~\bibnamefont
  {Jeong}}, \bibinfo {author} {\bibfnamefont {C.}~\bibnamefont {Jung}},
  \bibinfo {author} {\bibfnamefont {T.}~\bibnamefont {Kim}}, \emph {et~al.},\
  }\bibfield  {title} {\bibinfo {title} {{Using machine learning to improve
  multi-qubit state discrimination of trapped ions from uncertain EMCCD
  measurements}},\ }\href {https://doi.org/10.1364/OE.491301} {\bibfield
  {journal} {\bibinfo  {journal} {Opt. Express}\ }\textbf {\bibinfo {volume}
  {31}},\ \bibinfo {pages} {35113} (\bibinfo {year} {2023})}\BibitemShut
  {NoStop}%
\bibitem [{\citenamefont {Allcock}\ \emph {et~al.}(2021)\citenamefont
  {Allcock}, \citenamefont {Campbell}, \citenamefont {Chiaverini},
  \citenamefont {Chuang}, \citenamefont {Hudson}, \citenamefont {Moore},
  \citenamefont {Ransford}, \citenamefont {Roman}, \citenamefont {Sage},\ and\
  \citenamefont {Wineland}}]{allcock2021omg}%
  \BibitemOpen
  \bibfield  {author} {\bibinfo {author} {\bibfnamefont {D.}~\bibnamefont
  {Allcock}}, \bibinfo {author} {\bibfnamefont {W.}~\bibnamefont {Campbell}},
  \bibinfo {author} {\bibfnamefont {J.}~\bibnamefont {Chiaverini}}, \bibinfo
  {author} {\bibfnamefont {I.}~\bibnamefont {Chuang}}, \bibinfo {author}
  {\bibfnamefont {E.}~\bibnamefont {Hudson}}, \bibinfo {author} {\bibfnamefont
  {I.}~\bibnamefont {Moore}}, \bibinfo {author} {\bibfnamefont
  {A.}~\bibnamefont {Ransford}}, \bibinfo {author} {\bibfnamefont
  {C.}~\bibnamefont {Roman}}, \bibinfo {author} {\bibfnamefont
  {J.}~\bibnamefont {Sage}},\ and\ \bibinfo {author} {\bibfnamefont
  {D.}~\bibnamefont {Wineland}},\ }\bibfield  {title} {\bibinfo {title} {omg
  blueprint for trapped ion quantum computing with metastable states},\ }\href
  {https://doi.org/10.1063/5.0069544} {\bibfield  {journal} {\bibinfo
  {journal} {Appl. Phys. Lett.}\ }\textbf {\bibinfo {volume} {119}},\ \bibinfo
  {pages} {214002} (\bibinfo {year} {2021})}\BibitemShut {NoStop}%
\bibitem [{\citenamefont {Yang}\ \emph {et~al.}(2022)\citenamefont {Yang},
  \citenamefont {Ma}, \citenamefont {Wu}, \citenamefont {Wang}, \citenamefont
  {Cao}, \citenamefont {Guo}, \citenamefont {Huang}, \citenamefont {Feng},
  \citenamefont {Zhou},\ and\ \citenamefont {Duan}}]{yang2022realizing}%
  \BibitemOpen
  \bibfield  {author} {\bibinfo {author} {\bibfnamefont {H.-X.}\ \bibnamefont
  {Yang}}, \bibinfo {author} {\bibfnamefont {J.-Y.}\ \bibnamefont {Ma}},
  \bibinfo {author} {\bibfnamefont {Y.-K.}\ \bibnamefont {Wu}}, \bibinfo
  {author} {\bibfnamefont {Y.}~\bibnamefont {Wang}}, \bibinfo {author}
  {\bibfnamefont {M.-M.}\ \bibnamefont {Cao}}, \bibinfo {author} {\bibfnamefont
  {W.-X.}\ \bibnamefont {Guo}}, \bibinfo {author} {\bibfnamefont {Y.-Y.}\
  \bibnamefont {Huang}}, \bibinfo {author} {\bibfnamefont {L.}~\bibnamefont
  {Feng}}, \bibinfo {author} {\bibfnamefont {Z.-C.}\ \bibnamefont {Zhou}},\
  and\ \bibinfo {author} {\bibfnamefont {L.-M.}\ \bibnamefont {Duan}},\
  }\bibfield  {title} {\bibinfo {title} {Realizing coherently convertible
  dual-type qubits with the same ion species},\ }\href
  {https://doi.org/10.1038/s41567-022-01661-5} {\bibfield  {journal} {\bibinfo
  {journal} {Nat. Phys.}\ }\textbf {\bibinfo {volume} {18}},\ \bibinfo {pages}
  {1058} (\bibinfo {year} {2022})}\BibitemShut {NoStop}%
\bibitem [{\citenamefont {Shao}\ \emph {et~al.}(2023)\citenamefont {Shao},
  \citenamefont {Yue}, \citenamefont {Ma}, \citenamefont {Huang}, \citenamefont
  {Guan},\ and\ \citenamefont {Gao}}]{PhysRevResearch.5.023193}%
  \BibitemOpen
  \bibfield  {author} {\bibinfo {author} {\bibfnamefont {H.}~\bibnamefont
  {Shao}}, \bibinfo {author} {\bibfnamefont {H.}~\bibnamefont {Yue}}, \bibinfo
  {author} {\bibfnamefont {Z.}~\bibnamefont {Ma}}, \bibinfo {author}
  {\bibfnamefont {Y.}~\bibnamefont {Huang}}, \bibinfo {author} {\bibfnamefont
  {H.}~\bibnamefont {Guan}},\ and\ \bibinfo {author} {\bibfnamefont
  {K.}~\bibnamefont {Gao}},\ }\bibfield  {title} {\bibinfo {title} {{Precision
  determination of the $5d^{2}D_{3/2}$ state lifetime of single
  $^{174}\mathrm{Yb}^{+}$ ion}},\ }\href
  {https://doi.org/10.1103/PhysRevResearch.5.023193} {\bibfield  {journal}
  {\bibinfo  {journal} {Phys. Rev. Res.}\ }\textbf {\bibinfo {volume} {5}},\
  \bibinfo {pages} {023193} (\bibinfo {year} {2023})}\BibitemShut {NoStop}%
\bibitem [{\citenamefont {Acton}\ \emph {et~al.}(2005)\citenamefont {Acton},
  \citenamefont {Brickman}, \citenamefont {Haljan}, \citenamefont {Lee},
  \citenamefont {Deslauriers},\ and\ \citenamefont {Monroe}}]{acton2005near}%
  \BibitemOpen
  \bibfield  {author} {\bibinfo {author} {\bibfnamefont {M.}~\bibnamefont
  {Acton}}, \bibinfo {author} {\bibfnamefont {K.-A.}\ \bibnamefont {Brickman}},
  \bibinfo {author} {\bibfnamefont {P.}~\bibnamefont {Haljan}}, \bibinfo
  {author} {\bibfnamefont {P.}~\bibnamefont {Lee}}, \bibinfo {author}
  {\bibfnamefont {L.}~\bibnamefont {Deslauriers}},\ and\ \bibinfo {author}
  {\bibfnamefont {C.}~\bibnamefont {Monroe}},\ }\bibfield  {title} {\bibinfo
  {title} {Near-perfect simultaneous measurement of a qubit register},\
  }\href@noop {} {\bibfield  {journal} {\bibinfo  {journal} {arXiv preprint
  quant-ph/0511257}\ } (\bibinfo {year} {2005})}\BibitemShut {NoStop}%
\bibitem [{\citenamefont {Wang}\ \emph {et~al.}(2011)\citenamefont {Wang},
  \citenamefont {Ashhab},\ and\ \citenamefont {Nori}}]{PhysRevA.83.062317}%
  \BibitemOpen
  \bibfield  {author} {\bibinfo {author} {\bibfnamefont {H.}~\bibnamefont
  {Wang}}, \bibinfo {author} {\bibfnamefont {S.}~\bibnamefont {Ashhab}},\ and\
  \bibinfo {author} {\bibfnamefont {F.}~\bibnamefont {Nori}},\ }\bibfield
  {title} {\bibinfo {title} {Quantum algorithm for simulating the dynamics of
  an open quantum system},\ }\href {https://doi.org/10.1103/PhysRevA.83.062317}
  {\bibfield  {journal} {\bibinfo  {journal} {Phys. Rev. A}\ }\textbf {\bibinfo
  {volume} {83}},\ \bibinfo {pages} {062317} (\bibinfo {year}
  {2011})}\BibitemShut {NoStop}%
\bibitem [{\citenamefont {Han}\ \emph {et~al.}(2021)\citenamefont {Han},
  \citenamefont {Cai}, \citenamefont {Hu}, \citenamefont {Mu}, \citenamefont
  {Ma}, \citenamefont {Xu}, \citenamefont {Wang}, \citenamefont {Wang},
  \citenamefont {Song}, \citenamefont {Zou},\ and\ \citenamefont
  {Sun}}]{PhysRevLett.127.020504}%
  \BibitemOpen
  \bibfield  {author} {\bibinfo {author} {\bibfnamefont {J.}~\bibnamefont
  {Han}}, \bibinfo {author} {\bibfnamefont {W.}~\bibnamefont {Cai}}, \bibinfo
  {author} {\bibfnamefont {L.}~\bibnamefont {Hu}}, \bibinfo {author}
  {\bibfnamefont {X.}~\bibnamefont {Mu}}, \bibinfo {author} {\bibfnamefont
  {Y.}~\bibnamefont {Ma}}, \bibinfo {author} {\bibfnamefont {Y.}~\bibnamefont
  {Xu}}, \bibinfo {author} {\bibfnamefont {W.}~\bibnamefont {Wang}}, \bibinfo
  {author} {\bibfnamefont {H.}~\bibnamefont {Wang}}, \bibinfo {author}
  {\bibfnamefont {Y.~P.}\ \bibnamefont {Song}}, \bibinfo {author}
  {\bibfnamefont {C.-L.}\ \bibnamefont {Zou}},\ and\ \bibinfo {author}
  {\bibfnamefont {L.}~\bibnamefont {Sun}},\ }\bibfield  {title} {\bibinfo
  {title} {{Experimental Simulation of Open Quantum System Dynamics via
  Trotterization}},\ }\href {https://doi.org/10.1103/PhysRevLett.127.020504}
  {\bibfield  {journal} {\bibinfo  {journal} {Phys. Rev. Lett.}\ }\textbf
  {\bibinfo {volume} {127}},\ \bibinfo {pages} {020504} (\bibinfo {year}
  {2021})}\BibitemShut {NoStop}%
\bibitem [{\citenamefont {Barreiro}\ \emph {et~al.}(2011)\citenamefont
  {Barreiro}, \citenamefont {M{\"u}ller}, \citenamefont {Schindler},
  \citenamefont {Nigg}, \citenamefont {Monz}, \citenamefont {Chwalla},
  \citenamefont {Hennrich}, \citenamefont {Roos}, \citenamefont {Zoller},\ and\
  \citenamefont {Blatt}}]{barreiro2011open}%
  \BibitemOpen
  \bibfield  {author} {\bibinfo {author} {\bibfnamefont {J.~T.}\ \bibnamefont
  {Barreiro}}, \bibinfo {author} {\bibfnamefont {M.}~\bibnamefont
  {M{\"u}ller}}, \bibinfo {author} {\bibfnamefont {P.}~\bibnamefont
  {Schindler}}, \bibinfo {author} {\bibfnamefont {D.}~\bibnamefont {Nigg}},
  \bibinfo {author} {\bibfnamefont {T.}~\bibnamefont {Monz}}, \bibinfo {author}
  {\bibfnamefont {M.}~\bibnamefont {Chwalla}}, \bibinfo {author} {\bibfnamefont
  {M.}~\bibnamefont {Hennrich}}, \bibinfo {author} {\bibfnamefont {C.~F.}\
  \bibnamefont {Roos}}, \bibinfo {author} {\bibfnamefont {P.}~\bibnamefont
  {Zoller}},\ and\ \bibinfo {author} {\bibfnamefont {R.}~\bibnamefont
  {Blatt}},\ }\bibfield  {title} {\bibinfo {title} {An open-system quantum
  simulator with trapped ions},\ }\href {https://doi.org/10.1038/nature09801}
  {\bibfield  {journal} {\bibinfo  {journal} {Nature}\ }\textbf {\bibinfo
  {volume} {470}},\ \bibinfo {pages} {486} (\bibinfo {year}
  {2011})}\BibitemShut {NoStop}%
\bibitem [{\citenamefont {Mu\~noz Arias}\ \emph {et~al.}(2020)\citenamefont
  {Mu\~noz Arias}, \citenamefont {Poggi}, \citenamefont {Jessen},\ and\
  \citenamefont {Deutsch}}]{PhysRevLett.124.110503}%
  \BibitemOpen
  \bibfield  {author} {\bibinfo {author} {\bibfnamefont {M.~H.}\ \bibnamefont
  {Mu\~noz Arias}}, \bibinfo {author} {\bibfnamefont {P.~M.}\ \bibnamefont
  {Poggi}}, \bibinfo {author} {\bibfnamefont {P.~S.}\ \bibnamefont {Jessen}},\
  and\ \bibinfo {author} {\bibfnamefont {I.~H.}\ \bibnamefont {Deutsch}},\
  }\bibfield  {title} {\bibinfo {title} {{Simulating Nonlinear Dynamics of
  Collective Spins via Quantum Measurement and Feedback}},\ }\href
  {https://doi.org/10.1103/PhysRevLett.124.110503} {\bibfield  {journal}
  {\bibinfo  {journal} {Phys. Rev. Lett.}\ }\textbf {\bibinfo {volume} {124}},\
  \bibinfo {pages} {110503} (\bibinfo {year} {2020})}\BibitemShut {NoStop}%
\bibitem [{\citenamefont {Aksenov}\ \emph {et~al.}(2023)\citenamefont
  {Aksenov}, \citenamefont {Zalivako}, \citenamefont {Semerikov}, \citenamefont
  {Borisenko}, \citenamefont {Semenin}, \citenamefont {Sidorov}, \citenamefont
  {Fedorov}, \citenamefont {Khabarova},\ and\ \citenamefont
  {Kolachevsky}}]{PhysRevA.107.052612}%
  \BibitemOpen
  \bibfield  {author} {\bibinfo {author} {\bibfnamefont {M.~A.}\ \bibnamefont
  {Aksenov}}, \bibinfo {author} {\bibfnamefont {I.~V.}\ \bibnamefont
  {Zalivako}}, \bibinfo {author} {\bibfnamefont {I.~A.}\ \bibnamefont
  {Semerikov}}, \bibinfo {author} {\bibfnamefont {A.~S.}\ \bibnamefont
  {Borisenko}}, \bibinfo {author} {\bibfnamefont {N.~V.}\ \bibnamefont
  {Semenin}}, \bibinfo {author} {\bibfnamefont {P.~L.}\ \bibnamefont
  {Sidorov}}, \bibinfo {author} {\bibfnamefont {A.~K.}\ \bibnamefont
  {Fedorov}}, \bibinfo {author} {\bibfnamefont {K.~Y.}\ \bibnamefont
  {Khabarova}},\ and\ \bibinfo {author} {\bibfnamefont {N.~N.}\ \bibnamefont
  {Kolachevsky}},\ }\bibfield  {title} {\bibinfo {title} {{Realizing quantum
  gates with optically addressable $^{171}\mathrm{Yb}^{+}$ ion qudits}},\
  }\href {https://doi.org/10.1103/PhysRevA.107.052612} {\bibfield  {journal}
  {\bibinfo  {journal} {Phys. Rev. A}\ }\textbf {\bibinfo {volume} {107}},\
  \bibinfo {pages} {052612} (\bibinfo {year} {2023})}\BibitemShut {NoStop}%
\bibitem [{\citenamefont {Baldwin}\ \emph {et~al.}(2021)\citenamefont
  {Baldwin}, \citenamefont {Bjork}, \citenamefont {Foss-Feig}, \citenamefont
  {Gaebler}, \citenamefont {Hayes}, \citenamefont {Kokish}, \citenamefont
  {Langer}, \citenamefont {Sedlacek}, \citenamefont {Stack},\ and\
  \citenamefont {Vittorini}}]{PhysRevA.103.012603}%
  \BibitemOpen
  \bibfield  {author} {\bibinfo {author} {\bibfnamefont {C.~H.}\ \bibnamefont
  {Baldwin}}, \bibinfo {author} {\bibfnamefont {B.~J.}\ \bibnamefont {Bjork}},
  \bibinfo {author} {\bibfnamefont {M.}~\bibnamefont {Foss-Feig}}, \bibinfo
  {author} {\bibfnamefont {J.~P.}\ \bibnamefont {Gaebler}}, \bibinfo {author}
  {\bibfnamefont {D.}~\bibnamefont {Hayes}}, \bibinfo {author} {\bibfnamefont
  {M.~G.}\ \bibnamefont {Kokish}}, \bibinfo {author} {\bibfnamefont
  {C.}~\bibnamefont {Langer}}, \bibinfo {author} {\bibfnamefont {J.~A.}\
  \bibnamefont {Sedlacek}}, \bibinfo {author} {\bibfnamefont {D.}~\bibnamefont
  {Stack}},\ and\ \bibinfo {author} {\bibfnamefont {G.}~\bibnamefont
  {Vittorini}},\ }\bibfield  {title} {\bibinfo {title} {High-fidelity
  light-shift gate for clock-state qubits},\ }\href
  {https://doi.org/10.1103/PhysRevA.103.012603} {\bibfield  {journal} {\bibinfo
   {journal} {Phys. Rev. A}\ }\textbf {\bibinfo {volume} {103}},\ \bibinfo
  {pages} {012603} (\bibinfo {year} {2021})}\BibitemShut {NoStop}%
\bibitem [{\citenamefont {Hayes}\ \emph {et~al.}(2020)\citenamefont {Hayes},
  \citenamefont {Stack}, \citenamefont {Bjork}, \citenamefont {Potter},
  \citenamefont {Baldwin},\ and\ \citenamefont
  {Stutz}}]{PhysRevLett.124.170501}%
  \BibitemOpen
  \bibfield  {author} {\bibinfo {author} {\bibfnamefont {D.}~\bibnamefont
  {Hayes}}, \bibinfo {author} {\bibfnamefont {D.}~\bibnamefont {Stack}},
  \bibinfo {author} {\bibfnamefont {B.}~\bibnamefont {Bjork}}, \bibinfo
  {author} {\bibfnamefont {A.~C.}\ \bibnamefont {Potter}}, \bibinfo {author}
  {\bibfnamefont {C.~H.}\ \bibnamefont {Baldwin}},\ and\ \bibinfo {author}
  {\bibfnamefont {R.~P.}\ \bibnamefont {Stutz}},\ }\bibfield  {title} {\bibinfo
  {title} {{Eliminating Leakage Errors in Hyperfine Qubits}},\ }\href
  {https://doi.org/10.1103/PhysRevLett.124.170501} {\bibfield  {journal}
  {\bibinfo  {journal} {Phys. Rev. Lett.}\ }\textbf {\bibinfo {volume} {124}},\
  \bibinfo {pages} {170501} (\bibinfo {year} {2020})}\BibitemShut {NoStop}%
\bibitem [{\citenamefont {Ivory}\ \emph {et~al.}(2021)\citenamefont {Ivory},
  \citenamefont {Setzer}, \citenamefont {Karl}, \citenamefont {McGuinness},
  \citenamefont {DeRose}, \citenamefont {Blain}, \citenamefont {Stick},
  \citenamefont {Gehl},\ and\ \citenamefont {Parazzoli}}]{PhysRevX.11.041033}%
  \BibitemOpen
  \bibfield  {author} {\bibinfo {author} {\bibfnamefont {M.}~\bibnamefont
  {Ivory}}, \bibinfo {author} {\bibfnamefont {W.~J.}\ \bibnamefont {Setzer}},
  \bibinfo {author} {\bibfnamefont {N.}~\bibnamefont {Karl}}, \bibinfo {author}
  {\bibfnamefont {H.}~\bibnamefont {McGuinness}}, \bibinfo {author}
  {\bibfnamefont {C.}~\bibnamefont {DeRose}}, \bibinfo {author} {\bibfnamefont
  {M.}~\bibnamefont {Blain}}, \bibinfo {author} {\bibfnamefont
  {D.}~\bibnamefont {Stick}}, \bibinfo {author} {\bibfnamefont
  {M.}~\bibnamefont {Gehl}},\ and\ \bibinfo {author} {\bibfnamefont {L.~P.}\
  \bibnamefont {Parazzoli}},\ }\bibfield  {title} {\bibinfo {title}
  {{Integrated Optical Addressing of a Trapped Ytterbium Ion}},\ }\href
  {https://doi.org/10.1103/PhysRevX.11.041033} {\bibfield  {journal} {\bibinfo
  {journal} {Phys. Rev. X}\ }\textbf {\bibinfo {volume} {11}},\ \bibinfo
  {pages} {041033} (\bibinfo {year} {2021})}\BibitemShut {NoStop}%
\bibitem [{\citenamefont {Berkeland}\ and\ \citenamefont
  {Boshier}(2002)}]{PhysRevA.65.033413}%
  \BibitemOpen
  \bibfield  {author} {\bibinfo {author} {\bibfnamefont {D.~J.}\ \bibnamefont
  {Berkeland}}\ and\ \bibinfo {author} {\bibfnamefont {M.~G.}\ \bibnamefont
  {Boshier}},\ }\bibfield  {title} {\bibinfo {title} {Destabilization of dark
  states and optical spectroscopy in zeeman-degenerate atomic systems},\ }\href
  {https://doi.org/10.1103/PhysRevA.65.033413} {\bibfield  {journal} {\bibinfo
  {journal} {Phys. Rev. A}\ }\textbf {\bibinfo {volume} {65}},\ \bibinfo
  {pages} {033413} (\bibinfo {year} {2002})}\BibitemShut {NoStop}%
\end{thebibliography}%

\end{document}